\documentclass[journal]{IEEEtran}

\usepackage{amssymb}
\usepackage[cmex10]{amsmath}
\usepackage{stfloats}
\usepackage{graphicx}
\usepackage{subfigure}
\usepackage{tabularx}
\usepackage{epsfig,epsf,color,balance,cite}
\usepackage{verbatim}
\usepackage{url}
\usepackage{bm}

\usepackage{algorithm}
\usepackage{algorithmic}
\usepackage{enumitem}

\begin{document}

\title{Secure Semantic Communications: Fundamentals and Challenges}
\author{Zhaohui Yang, Mingzhe Chen, Gaolei Li, Yang Yang, and Zhaoyang Zhang
\thanks{Zhaohui Yang and Zhaoyang Zhang are with the College of Information Science and Electronic Engineering, Zhejiang University, Hangzhou 310027, China, and Zhejiang Provincial Key Lab of Information Processing, Communication and Networking (IPCAN), Hangzhou 310007, China. Zhaohui Yang is also with Zhejiang Lab, Hangzhou 31121, China. (e-mails: yang\_zhaohui@zju.edu.cn and ning\_ming@zju.edu.cn).}
\thanks{Gaolei Li is with the School of Electronic Information and Electrical Engineering, Shanghai Jiao Tong University, Shanghai, China. (e-mail: gaolei\_li@sjtu.edu.cn) }
\thanks{Yang Yang is with the School of Information and Communication Engineering, BUPT.  (e-mail: yangyang01@bupt.edu.cn) }
\thanks{Mingzhe Chen is with the Department of Electrical and Computer Engineering and Institute for Data Science and Computing, University of Miami, Coral Gables, FL 33146 USA. (e-mail: mingzhe.chen@miami.edu)}
 }
\maketitle

\begin{abstract}
Semantic communication allows the receiver to know the intention instead of the bit information itself, which is an emerging technique to support  real-time human-machine and machine-to-machine interactions for future wireless communications. 
In semantic communications, both transmitter and receiver
share some common knowledge, which can be used to extract small-size information at the transmitter and recover the original information at the receiver. 
Due to different design purposes, security issues in semantic communications have two unique features compared to standard bit-wise communications. First, an attacker
in semantic communications considers not only the amount of
stolen data but also the meanings of stolen data. Second, an attacker in semantic communication
systems can attack not only semantic information transmission
as done in standard communication systems but also attacks
machine learning (ML) models used for semantic information
extraction since most of semantic information is generated
using ML based methods. Due to these unique features, 
in this paper, we present  an overview on the fundamentals and  key challenges in the design of secure semantic communication.
We first provide various methods to define and extract semantic information. 
Then, we focus on secure semantic communication techniques in two areas: information security and semantic ML model security. For each area, we identify the main problems and challenges. Then, we will provide a comprehensive treatment of these problems. In a nutshell,
this article provides a holistic set of guidelines on how to design secure semantic communication systems over real-world wireless communication networks.
\end{abstract}
\begin{IEEEkeywords}
Secure semantic communication, information security, semantic ML model security.
\end{IEEEkeywords}

\IEEEpeerreviewmaketitle

\section{Introduction}
 The development of smartphone processors enable edge devices (i.e., mobile phones) to generate and process large-scale image, video, and immersive extended reality data, which will significantly increase network congestion \cite{saad2019vision}. Therefore, it is necessary to design novel communication techniques to support such large data-sized data transmission and processing. Current research \cite{saad2019vision} studied the use of machine learning (ML) tools, reflecting intelligent surface (RIS), millimeter wave (mmWave), and edge computing to improve network performance. However, the performance of a network that exploits these techniques will be limited by the Shannon capacity since most of these techniques are trying to maximize edge devices' data rates so as to reach the Shannon capacity limit \cite{shannon1948mathematical}. In consequence, it is necessary to design novel communication techniques that further improve network performance beyond the Shannon capacity limit.

 Semantic communication technique is a promising method to overcome the Shannon capacity limit, which enables an edge device to extract the meaning of large-sized data, called semantic information hereinafter, and transmit only the semantic information to the receiver instead of transmitting the entire data \cite{qin2021semantic}. Therefore, compared to current works that only focus on the maximization of devices' data rates, the purpose of semantic communications is not only to maximize each device's data rate but also maximize the meanings that the transmitted data can carry. Since semantic communication is still in its infancy, it faces many challenges such as semantic information definition, semantic information extraction, semantic communication measurement, security issues, and resilience.   

 Recently, a number of surveys and tutorials related to semantic communications appeared in \cite{qin2021semantic,9663101,yang2022semantic,9530497}. In particular, the authors in \cite{qin2021semantic,9663101,yang2022semantic} provided a comprehensive tutorial on the use of information theory for semantic information representation and semantic communication metric design. The authors in \cite{9530497} introduced an edge intelligence based semantic communication framework and present its implementation challenges. However, none of these existing surveys and tutorials \cite{qin2021semantic,9663101,yang2022semantic,9530497} introduced the security problems in semantic communications. Compared to attackers that only consider the amount of stolen data in standard communication systems, attackers in semantic communications have two unique features. First, an attacker in semantic communications considers not only the amount of stolen data but also the meanings of stolen data. For example, if one attacker steals a large amount of data from a user but does not obtain the target content/meanings from the stolen data, we will consider that the attacker does not attack the user successfully. Second, an attacker in semantic communication systems can attack not only semantic information transmission as done in standard communication systems but also attacks 
ML models used for semantic information extraction since most of semantic information is generated using ML based methods. Due to these unique features of attackers in semantic communications, it is necessary to provide an introduction on the fundamentals and challenges of implementing secure semantic communications.

 In this paper, we introduce fundamentals, solutions, and challenges of designing secure semantic communication systems. In this context, we first introduce basic semantic communication process. Then, we overview four methods to define semantic information: a) autoencoder, b) information bottleneck (IB), c) knowledge graph, and, d) probability graph, and summarize their advantages, drawbacks, and applications. We then introduce how attackers can attack semantic information transmission and extraction, and explain the methods to defense these attacks from the point views of both information security and ML security.
\section{Fundamentals of Semantic Communications}
In this section, we first the process of semantic communications. Then, we 
introduce four methods to model the semantic information extracted from original data and explain their differences, advantages, disadvantages, and applications. 

\begin{figure}[t]
  \centering
  \includegraphics[width=0.5\textwidth]{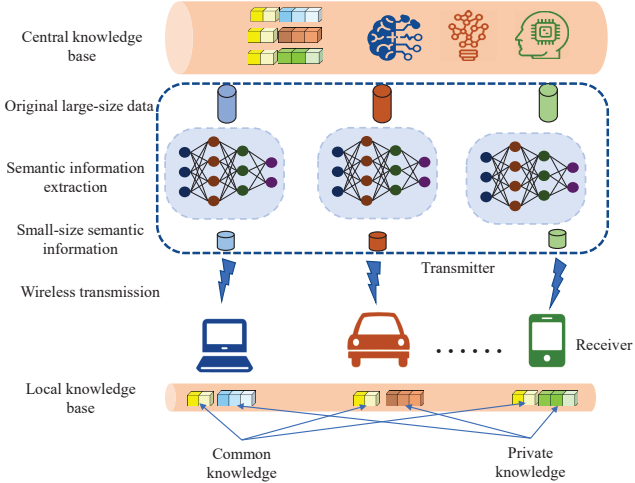}
  \caption{Illustration of the basic structures of a semantic communication system.}
\label{fig:OverallSemantic1}
\end{figure}

\subsection{Semantic Communication Process}
The overall semantic communication process mainly includes three stages. 
In the first stage,   the transmitter utilizes ML tool to extract the small-size semantic information from the original large-size data based on the central knowledge base, as shown in Fig.~\ref{fig:OverallSemantic1}.
Then, the semantic information is transmitted over wireless link in the second stage. 
In the third stage, the receiver recovers the intended meaning behind the semantic information based on its own local knowledge base, which includes both common knowledge and private knowledge. 
\subsection{Semantic Information Construction}

\subsubsection{Autoencoder}
Semantic communication transmits semantic messages, which refer to a sequence of well-formed symbols learned from the “meaning” underlying source. Correspondingly, the receiver aims at fully understanding the ``meaning'' of the encoded semantic symbols.
Therefore, effectively extracting the semantics of the source while ignoring the redundant information plays a key role in semantic communications. 
Due to the powerful representation and learning capability, neural networks are typically employed to extract semantics from the source. In particular, autoencoder is a type of neural network used to learn efficient representation for high-dimensional data, which can extract the most important information and is thus particularly suitable for semantic communications. 
In particular, autoencoder consists of an encoder and a decoder. The encoder outputs encoded symbols with much fewer dimensions compared to the source data, since it only reserves the key information while discards the insignificant parts of the data. Then, the decoder is used to recover the original data from the low-dimensional symbols. 

Using autoencoder to extract the semantic information has following advantages. First, autoencoder can be implemented based on various types of neural networks such as convolutional neural network, transformer, and fully-connected neural networks. Therefore, it is applicable to semantic communications of different source data including text, images, videos, and multi-modal data. In addition, since the output of the encoder has much less dimensions, the transmission efficiency of semantic communications can be significantly improved. Moreover, autoencoders can be trained in an self-supervised way. However, there are also some key challenges of autoencoders. In particular, even though the coding generated by the autoencoder can be efficiently understood by machines, it is incomprehensible for humans, which seems to be contradicted to the principle of semantic communications in a certain way. Table \ref{tab:trajectory} summarizes the advantages, disadvantages, and applications of using autoencoder to model semantic information.

\begin{figure*}[t]
  \centering
  \includegraphics[width=1\textwidth]{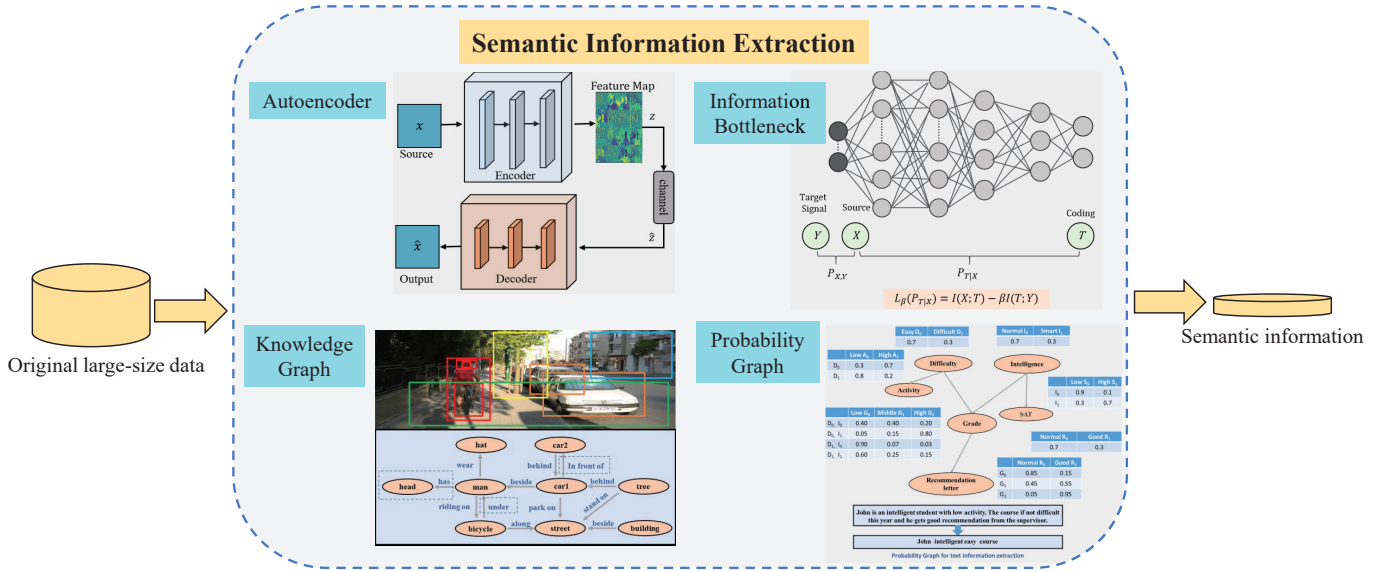}
  \caption{Illustration of four types of semantic information extraction.}
\label{fig：Semanticextraction}
\end{figure*}
\subsubsection{Information  Bottleneck}
Since semantic communications aim to only reserve the semantics, the essence of semantic communications is a lossy compression problem. To solve this type of problems, Claude Shannon has proposed fundamental theory, i.e. rate-distortion theory, which solves the optimal trade-off between compression and fidelity \cite{Shannon1959Coding}. 
In particular, rate-distortion theory aims to minimize the required rate under a given distortion, which can be used to guide the training process of semantic communication systems. However, one problem of rate-distortion theory is that it needs to choose a specific distortion function in advance, which will further determine the extracted semantics. However, the choice of the distortion function is not part of the theory.   

To tackle this issue, information bottleneck (IB) principle was proposed from the perspective of information theory, which can be deemed as a generalization of rate-distortion theory \cite{Goldfeld2020The}. On the one hand, the distortion in IB principle is measured by the mutual information between the encoded semantic symbols and a target variable. In semantic communications, the target variable varies with the applications. For instance, for an image classification task, the target variable is the label of the source image, since we try to correctly classify the source image. On the other hand, the rate in IB principle is characterized  by the mutual information between the source and the encoded symbols, which indicates the number of bits the encoded symbols used to represent the source. 

The advantage of IB is that it provides a specific theoretical bound for minimizing the rate under given distortion. However, in practice, the joint and marginal distributions of information bits are challenging to obtain and, thus, the original IB cannot be directly used to guide the training process of semantic communications. Table \ref{tab:trajectory} summarizes the advantages, disadvantages, and applications of using information bottleneck to model semantic information.

\subsubsection{Knowledge Graph} Since semantic information represented by the output of autoencoder is incomprehensible for humans and does not have any physical meaning, next, we introduce the use of knowledge graph for semantic information representation. Since a knowledge graph consists of nodes and edges, we use nodes to represent an object or a concept in the original data and edges are used to represent the relations between each pair of nodes.  Each pair of nodes and their relations are defined by a triple. Hence, a semantic information modeled by a knowledge graph consists of multiple triples. Different from other graph models, triples in knowledge graph are determined by both original data that a user wants to transmit and the knowledge that this user has to understand the original data. Hence, the semantic information extracted from the same original data by different users with different knowledge may be different.

Using knowledge graph to model semantic information has several key advantages. First, the extracted semantic information is comprehensible for
humans. Hence, a receiver may not need to recover the original data when it receives a semantic information since the semantic information represents the similar meanings of original data. Second, one can manage the data size of semantic information that consists of several triples according to network conditions and resources. In particular, when the network resources (i.e., bandwidth) are limited, one can limit the number of triples in the semantic information to meet communication service requirements (i.e., delay). However, exploiting knowledge graphs for semantic information also faces several challenges. First, all triples in a semantic information are extracted by neural network based methods. Therefore, the complexity and time of training these neural networks must be considered and reduced when using knowledge graph for energy limited devices (i.e., Internet-of-things devices). Second, most of current researches assume that all users have the same knowledge for triple extractions which may not be practical. Hence, it is necessary to design novel methods to model and generate unique knowledge library for each user. Table \ref{tab:trajectory} summarizes the advantages, disadvantages, and applications of using knowledge graph to model semantic information.   

\subsubsection{Probability Graph}
The directional probability graph can also be used to characterize the inherent information of the transmitted information \cite{griffiths2019probabilistic}. 
In the directional probability graph, each vertex represents the semantic entity and the edge stands for the probability of connection between these two vertexes. Since multiple vertexes with high connection probabilities among each other can be fused into a single vertex, the new generated vertex can contain higher level semantic information than the original vertexes. Probability graph shows the probabilities among different entities, which can be used to extract the corn semantic information with overall high probability in the probability  graph. 

There are some advantages of extracting semantic information with probability graph. First, different levels of semantic information can be generated with using probability with combing highly-rated low-level semantic entity into a high-level semantic entity. Second, the probability graph can be used to predict the incoming information of the receiver. Through prediction and inference, the receiver side can adapt its actions in advance. 
However, there are still some challenges using probability graph.  Since the probability graph can be  learned with neural network and multi-level semantic information extractions needs additional computation, the complexity of constructing multi-level semantic information is high. Besides, to ensure that the receiver can precisely predict the future information of the transmitter, both the transmitter  and receiver needs to share some highly related common knowledge.  
Table \ref{tab:trajectory} summarizes the advantages, disadvantages, and applications of using probability graph to model semantic information.
As a result, 
Fig, \ref{fig：Semanticextraction} illustrates the semantic extraction processes of above four methods.
		

\begin{table*}[htbp]
\centering
\caption{Various Semantic Information Construction Methods}
\label{tab:trajectory}
\begin{tabular}{|p{3.0cm}|p{5.0cm}|p{5.0cm}|p{3cm}|}
\hline
Classifications&  Advantages & Disadvantages &    Applications \\
\hline
\vspace{2em}
   Autoencoder 
 &   \begin{itemize}[leftmargin=*]
    \item Autoecnoder is applicable to data of different modalities
    \item Autoencoder can be trained in an self-supervised way
\end{itemize} &  
   \begin{itemize}[leftmargin=*]
    \item The extracted semantic information is incomprehensible for humans
\end{itemize} & 
  \begin{itemize}[leftmargin=*]
    \item Deep joint source-channel coding
    \item Multi-modal semantic transmission
\end{itemize}
\\
\hline
\vspace{1.5em}
Information  Bottleneck  & \begin{itemize}[leftmargin=*]
    \item IB provides a specific theoretical bound for minimizing the rate under given distortion
\end{itemize}   &  
\begin{itemize}[leftmargin=*]
    \item The mutual information is challenging to be obtained
    \item The extracted semantic information is incomprehensible for humans
\end{itemize}
& 
\begin{itemize}[leftmargin=*]
    \item Task-oriented semantic transmission
\end{itemize}
\\
 \hline
 \vspace{2em}
  Knowledge Graph &   \begin{itemize}[leftmargin=*]
    \item Modeled semantic information has physical meanings
    \item Receivers do not need to recover original data
\end{itemize}  &
   \begin{itemize}[leftmargin=*]
    \item Senders need to generate knowledge library for ML model training
    \item Token selection is implemented by complex neural networks
\end{itemize}
  &
   \begin{itemize}[leftmargin=*]
    \item Machine to machine communications
    \item Human to machine communications
\end{itemize}\\
 \hline
 \vspace{2em}
 Probability Graph & 
   \begin{itemize}[leftmargin=*]
    \item Probability graph can be utilized to conduct inference 
    \item Different levels of semantic information can be extracted
\end{itemize} &
   \begin{itemize}[leftmargin=*]
    \item High computation complexity 
   \item Both the transmitter and receiver need to
share some common knowledge

\end{itemize}
& 
   \begin{itemize}[leftmargin=*]
    \item Multi-level semantic information
\item Information prediction and inference
\end{itemize}
\\
 \hline
 \end{tabular}
\end{table*}





\section{Information Security}
Next, we introduce security issues in semantic information transmission. In particular, we discuss the security issues in information bottleneck, convert communications, and physical layers and summarize the challenges of implementing secure semantic communications in these scenarios.
\subsection{Information Bottleneck Security}
In this subsection, we introduce the use of IB for secure semantic communications. As previously mentioned, IB utilizes the mutual information between the source and the encoded symbols, and that between the encoded symbols and the target variable to represent the rate and distortion of a semantic communication system respectively. Following this principle, one can further extend IB to secure semantic communications. In particular, when an attacker exists who tries to detect the sensitive semantic information of the legitimate users during semantic communications, one can lower the sensitive semantic information leakage probability by minimizing the mutual information between the sensitive information and encoded semantic symbols. In this way, IB simultaneously considers rate, distortion, and security of the semantic communications and, thus, its target becomes to minimize the rate under given distortion and information leakage dangers.   

In addition, IB may also be used to analyze the training process of a secure semantic communication system. In particular, we can employ IB as the loss function to train a semantic communication system. During the training, we can estimate the values of the three mutual information terms, and thus the performance trade-off process of the rate, distortion, and security in each stage of the training can be analyzed. In this way, we can better understand the training process of the secure semantic communication systems, and further choose proper hyberparameters accordingly to optimize the performance of the systems.

However, applying IB to instruct secure semantic communications still has some key challenges. In particular, similar to the origianl form of IB, the mutual information is challenging to be calculated since the joint and marginal distributions of the encoded symbols and the sensitive information are typically unknown. Therefore, tight, trainable bounds for the mutual information terms are needed to effectively train the semantic communications. 

\subsection{Convert Semantic Communications}
In this subsection, we introduce the use of convert communications for semantic information transmission. In semantic communications, we consider not only the amount of data that is detected by an attacker but also the detected meanings of the original data. Therefore, we can use two methods to protect a sender's data privacy. First, we can use traditional convert communication techniques to protect a sender's data privacy. In particular, one can transmit interference signals to the user that the attacker wants to attack so as to protect the user's data privacy. Second, one can modify the triples in semantic information so as to avoid carrying the triples that an attacker wants to detect. We can also jointly consider the use of traditional convert communication methods and triple selection methods to protect users' data security.

However, using convert communications for secure semantic communications also faces several challenges. First, the sender needs to estimate the triples that attackers want to detect as well as the triples that the receiver wants to receive since the sender will not communicate with the attackers and receiver before semantic information transmission. Second, it is necessary to analyze energy consumption, data transmission delay, and other costs of implementing these two data protection methods. To this end, one can determine to use traditional convert communication methods or triple selection methods for different network conditions. For example, when the sender has limited transmit power for semantic information transmission, we may not be able to use traditional convert communication methods to protect user's data security since the receiver may not be able to detect the original signal of the sender. Finally, investigating the cooperation between the sender and receiver to further improve data security is another interesting research direction.



%
\subsection{Physical Layer Security}
The physical layer security \cite{shiu2011physical,mukherjee2014principles} of semantic communication includes two aspects:  common knowledge base security and semantic information security.
\begin{figure*}[t]
  \centering
  \includegraphics[width=1\textwidth]{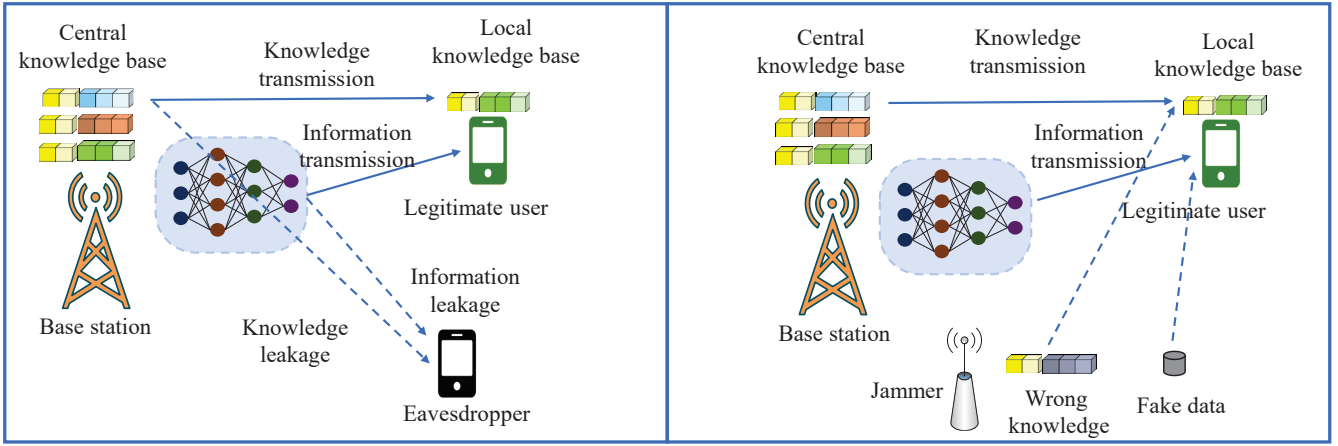}
  \caption{Illustration of the physical layer security in semantic communication systems with eavesdropper
 and jammer.}
\label{fig:PLS}
\end{figure*}

\subsubsection{Knowledge Base Security}
The common knowledge base is shared among the transmitter and the receiver, which helps the transmitter to  extract semantic information and allows the receiver to recover the behind  information of the received bits. 
As a result, it is of importance to protect the common knowledge base. As shown in Fig. \ref{fig:PLS}, the physical layer security of semantic communication includes two cases. For the eavesdropping case, the knowledge of the transmitter can be leakaged. For the jamming case, a jammer can send wrong knowledge, which can lead the legitimate user to recover the wrong message from the received semantic information. To protect the knowledge security, beamforming design and secure key distribution method can be applied. 
\subsubsection{Semantic Information Security}
The semantic information is based on the shared knowledge between the transmitter and the receiver,  which indicates that the semantic information relates to the conditional information. 
According to the construction procedures of semantic information, in additional to conventional interference and noise, there also exist semantic interference and noise.
The semantic interference includes the  ambiguity of the information.
For example, one word or phrase can have different meanings in various scenarios. With limited background information, it is hard to distinguish the real meaning of the specific semantic information, which causes semantic interference.  
The semantic noise is introduced since the original information is mapped into the semantic space, which can introduce additional noise since multiple information can be mapped into the same or similar semantic information. 
Thus, the secure semantic information rate should take into the semantic interference and noise into consideration , which calls for new design to protect semantic information security. One possible way is a joint learning and communication design  \cite{chen2021a}   to ensure the security of semantic communication.

\section{Semantic ML Security}
As mentioned in Section II, many ML techniques such as knowledge graph, encoder/decoder, and deep neural networks are perceived as important enablers for constructing basic components of semantic communication systems. In this section, we denote these enablers as ``semantic ML". It is of great significance to predict, model, and analyze the emerging security risks of semantic ML in advance for the development and popularization of future semantic communication systems. For example, all semantic features uploaded to the knowledge base should have sophisticated censorship mechanisms to prevent from being modified by malicious users. In this section, we firstly analyze the vulnerabilities of semantic ML and known threats. And then, we also discuss promising countermeasures against these threats. Finally, we envision the potential of trustworthy and explainable technologies for constructing secure semantic ML models for semantic communication systems.

\subsection{Security Risks}
\begin{figure}[t]
  \centering
  \includegraphics[width=0.46\textwidth]{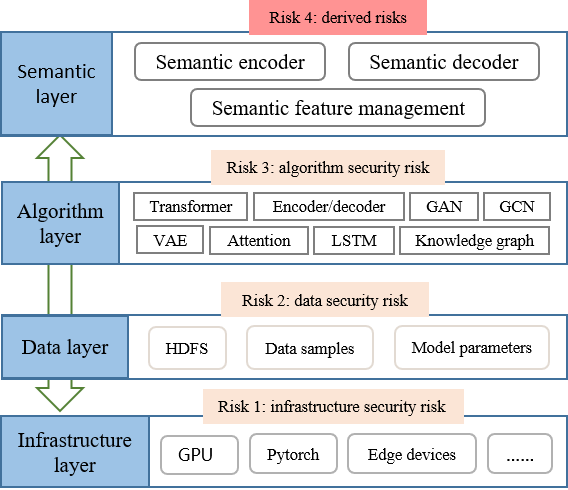}
  \caption{Hierarchical components of semantic ML and its security risks.}
\label{fig:AT-CM}
\end{figure}

Main security risks of semantic ML can be summarized as endogenous risks and derived risks, as illustrated in Fig. ~\ref{fig:AT-CM}. 

\subsubsection{Endogenous Risks}
For semantic ML, data is an important driver, algorithm is the core technology, and infrastructure is the underlying support.
Endogenous risks may result in data security threats, algorithm security threats, and infrastructure security threats.

\textbf{Risk 1: data security risk}. Known data security risks \cite{9743317} include privacy leakage, data poisoning, data pollution, data forgery, etc. In semantic communication systems, we may use ML to extract semantic features, generate knowledge base, and parse transmitted semantic features as original data. During the model training process of each semantic communication node, if the opponent tampers some original data samples by poisoning, polluting, etc.), the accuracy of data transmission will decrease significantly. Moreover, if a particular backdoor is injected into the ML models of semantic communication, that is, forcing the ML models to be extremely sensitive to a specific trigger (content with specific semantics), then the entire semantic communication system will be vulnerable to malicious manipulation. Besides, with the emerging of model inversion, gradient leakage and membership inference, both black-box and white-box data reconstruction  may threaten the practicality of ML in semantic communication systems, which is very difficult to prevent.

\textbf{Risk 2: Algorithm security}. Black-box and fairness are the two major sources of algorithm security risks. On  one hand, the ML algorithm based on neural networks extracts the features of the input data through a complex calculation process, and then classifies it into the calibrated class, but existing scientific knowledge and principles cannot give a reasonable explanation for the classification results, which is the fundamental reason why attacks are difficult to be effectively defended against. For example, it is difficult to locate where there is a fault when the receiver in a neural network-based semantic communication system outputs wrong results. On the other hand, in the process of extracting knowledge from massive data, algorithms such as sorting, classifying, associating, filtering, and attention mechanism are usually used to process the data items. If the algorithms exhibit unfair outputs under different gender or ethnicity settings, information transmission using semantic communication systems will suffer from complex social problems. 

\textbf{Risk 3: Infrastructure security}. To train semantic ML models and provide intelligent communication services, it is necessary to construct various development libraries, connect heterogeneous computing devices and build cloud platforms. Once vulnerabilities in these semantic ML infrastructures are exploited by attackers, the semantic ML models generated through these infrastructures may exhibit abnormal behaviors. For example, malwares on Tensorflow and PyTorch may modify model structures, while some hackers will try to illegally occupy the computing resources (for example GPU, CPU, and virtual machines) of deep learning nodes for mining. 

\subsubsection{Derived Risks}
In addition to existing endogenous risks, deploying ML models in semantic communication also face many derived risks due to the openness of real application scenarios. Derived risks mainly contain semantic adversarial samples and semantic backdoors. For example, the vision transformer that can be used to construct real-time semantic communications is vulnerable to adversarial examples \cite{9710333}. Besides, derived risks also contain man-in-the-middle attacks, DDoS attacks, and signal interference attacks, because ML models on the senders and receivers of semantic communications need cross-domain training.

\subsection{Countermeasures}
\begin{figure*}[ht]
  \centering
  \includegraphics[width=0.83\textwidth]{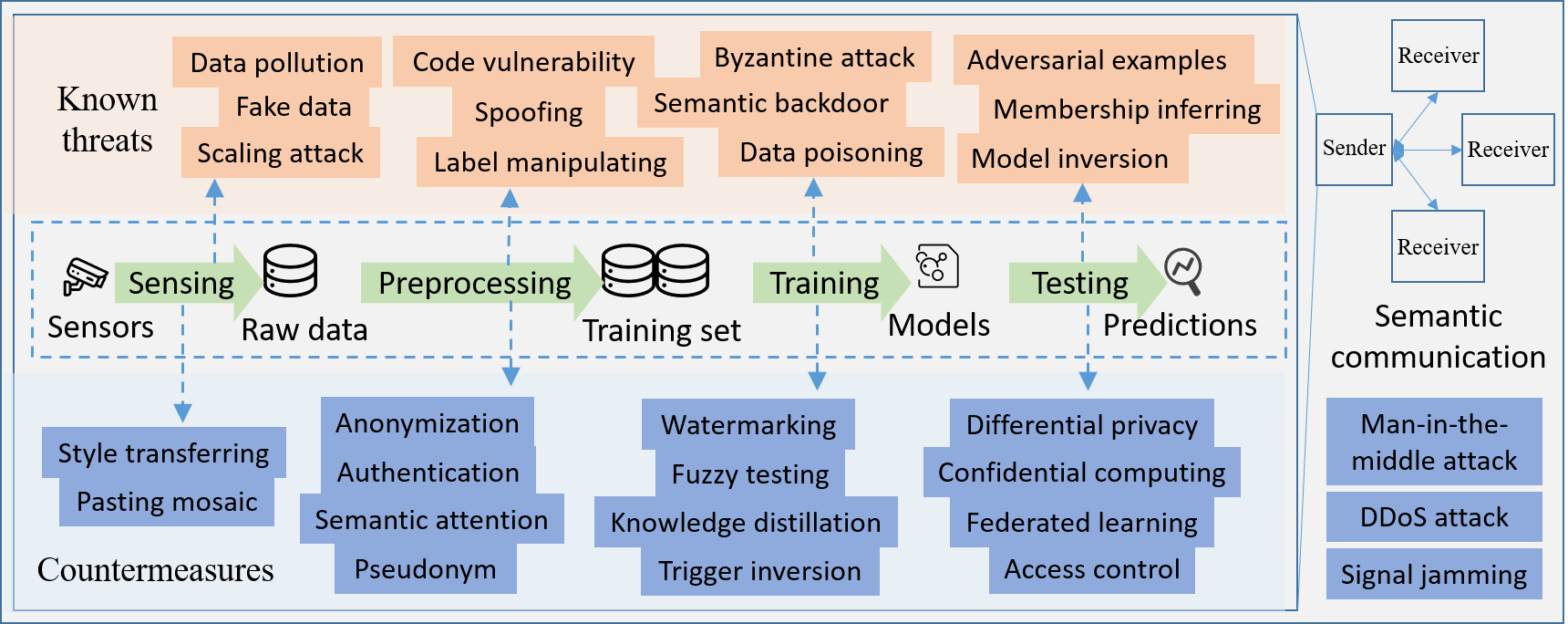}
  \caption{Illustration of known threats and countermeasures in the ML lifecycle. Each intelligent agent in the semantic communication network that is configured with at least one ML model may suffer from various attacks. When senders and receivers need to train the encoder/decoder networks jointly, the semantic communication can also suffer from security threats from transmission channels.}
\label{fig:AT-CM2}
\end{figure*}

To bring out how to attack a ML model with above security risks in semantic communication systems, Fig.~\ref{fig:AT-CM2} illustrates most of known attacks. According to features of known attacks, corresponding countermeasures can be mainly divided into three branches: 1) Anti-poisoning; 2) Improving the robustness to adversarial examples; 3) Preventing privacy leakage. As for anti-poisoning methods, there are three different designing paradigms, including a) removing poisoned data, b) smoothing abnormal activation values, and c) erasing hidden backdoors. Both a) and b) require to access the model training process, which is often unpractical for service providers. Trigger inversion and data-free knowledge distillation acting as black-box defence techniques are being considered increasingly. 

For model robustness, the most popular countermeasures are adversarial training and differential privacy. However, adversarial training often needs more data samples and the differential privacy may reduce model accuracy, so that combining model interpretability to explore the causes of adversarial samples becomes current hotspots. To prevent an attacker from recovering the original data and semantics from the ML model, privacy-preserving frameworks such as confidential computing, differential privacy \cite{9158374}, and federated learning are being studied to satisfy individual requirements in different scenarios. In different application scenarios, users need to choose appropriated privacy preserving methods based on their own individual requirements. Besides, many artifical intelligence service providers attach great importance to model watermarking and authentication techniques in order to secure their intellectual property rights.


\section{Conclusion} 
In this paper, we have provided four kinds of methods to represent the semantic information. Based on the construction of semantic information, we pointed out that the main securities in semantic information followed into two aspects, information security and ML security. 
For each aspect, we have pointed out the main problems and the corresponding treatments. Future directions include the joint communication and computation design for performance analysis and optimization of secure semantic communication.

\bibliographystyle{IEEEtran}
\bibliography{IEEEabrv,MC,Ref}

\begin{thebibliography}{10}
\providecommand{\url}[1]{#1}
\csname url@samestyle\endcsname
\providecommand{\newblock}{\relax}
\providecommand{\bibinfo}[2]{#2}
\providecommand{\BIBentrySTDinterwordspacing}{\spaceskip=0pt\relax}
\providecommand{\BIBentryALTinterwordstretchfactor}{4}
\providecommand{\BIBentryALTinterwordspacing}{\spaceskip=\fontdimen2\font plus
\BIBentryALTinterwordstretchfactor\fontdimen3\font minus
  \fontdimen4\font\relax}
\providecommand{\BIBforeignlanguage}[2]{{%
\expandafter\ifx\csname l@#1\endcsname\relax
\typeout{** WARNING: IEEEtran.bst: No hyphenation pattern has been}%
\typeout{** loaded for the language `#1'. Using the pattern for}%
\typeout{** the default language instead.}%
\else
\language=\csname l@#1\endcsname
\fi
#2}}
\providecommand{\BIBdecl}{\relax}
\BIBdecl

\bibitem{saad2019vision}
W.~{Saad}, M.~{Bennis}, and M.~{Chen}, ``A vision of {6G} wireless systems:
  {A}pplications, trends, technologies, and open research problems,''
  \emph{IEEE Network}, vol.~34, no.~3, pp. 134--142, May/June 2020.

\bibitem{shannon1948mathematical}
C.~E. Shannon and W.~Weaver, \emph{The Mathematical Theory of
  Communication}.\hskip 1em plus 0.5em minus 0.4em\relax The University of
  Illinois Press, 1949.

\bibitem{qin2021semantic}
Z.~Qin, X.~Tao, J.~Lu, and G.~Y. Li, ``Semantic communications: {P}rinciples
  and challenges,'' \emph{arXiv preprint arXiv:2201.01389}, 2021.

\bibitem{9663101}
Q.~Lan, D.~Wen, Z.~Zhang, Q.~Zeng, X.~Chen, P.~Popovski, and K.~Huang, ``What
  is semantic communication? a view on conveying meaning in the era of machine
  intelligence,'' \emph{Journal of Communications and Information Networks},
  vol.~6, no.~4, pp. 336--371, 2021.

\bibitem{yang2022semantic}
W.~Yang, H.~Du, Z.~Liew, W.~Y.~B. Lim, Z.~Xiong, D.~Niyato, X.~Chi, X.~S. Shen,
  and C.~Miao, ``Semantic communications for {6G} future internet:
  {F}undamentals, applications, and challenges,'' \emph{arXiv preprint
  arXiv:2207.00427}, 2022.

\bibitem{9530497}
G.~Shi, Y.~Xiao, Y.~Li, and X.~Xie, ``From semantic communication to
  semantic-aware networking: {M}odel, architecture, and open problems,''
  \emph{IEEE Communications Magazine}, vol.~59, no.~8, pp. 44--50, Aug. 2021.

\bibitem{Shannon1959Coding}
C.~E. {Shannon}, ``Coding theorems for a discrete source with a fidelity
  criterion,'' \emph{IRE Conv. Rec.}, vol.~7, no.~9, pp. 56--83, 1959.

\bibitem{Goldfeld2020The}
Z.~{Goldfeld} and Y.~{Polyanskiy}, ``The information bottleneck problem and its
  applications in machine learning,'' \emph{IEEE J. Sel. Areas Commun.},
  vol.~1, no.~1, pp. 19--38, 2020.

\bibitem{griffiths2019probabilistic}
T.~L. Griffiths and M.~Steyvers, ``A probabilistic approach to semantic
  representation,'' in \emph{Pro. Twenty-Fourth Annual Conf. Cognitive Science
  Society}.\hskip 1em plus 0.5em minus 0.4em\relax Routledge, 2019, pp.
  381--386.

\bibitem{shiu2011physical}
Y.-S. Shiu, S.~Y. Chang, H.-C. Wu, S.~C.-H. Huang, and H.-H. Chen, ``Physical
  layer security in wireless networks: {A} tutorial,'' \emph{IEEE Wireless
  Commun.}, vol.~18, no.~2, pp. 66--74, 2011.

\bibitem{mukherjee2014principles}
A.~Mukherjee, S.~A.~A. Fakoorian, J.~Huang, and A.~L. Swindlehurst,
  ``Principles of physical layer security in multiuser wireless networks: {A}
  survey,'' \emph{IEEE Commun. Surveys \& Tutor.}, vol.~16, no.~3, pp.
  1550--1573, 2014.

\bibitem{chen2021a}
M.~Chen, Z.~Yang, W.~Saad, C.~Yin, H.~V. Poor, and S.~Cui, ``A joint learning
  and communications framework for federated learning over wireless networks,''
  \emph{IEEE Trans. Wireless Commun.}, vol.~20, no.~1, pp. 269--283, Jan. 2021.

\bibitem{9743317}
M.~Goldblum, D.~Tsipras, C.~Xie, X.~Chen, A.~Schwarzschild, D.~Song, A.~Madry,
  B.~Li, and T.~Goldstein, ``Dataset security for machine learning: Data
  poisoning, backdoor attacks, and defenses,'' \emph{IEEE Trans. Pattern
  Analysis and Machine Intelligence}, pp. 1--1, 2022.

\bibitem{9710333}
K.~Mahmood, R.~Mahmood, and M.~van Dijk, ``On the robustness of vision
  transformers to adversarial examples,'' in \emph{Proc. IEEE/CVF Int. Conf.
  Computer Vision (ICCV)}, 2021, pp. 7818--7827.

\bibitem{9158374}
T.~Zhu, D.~Ye, W.~Wang, W.~Zhou, and P.~S. Yu, ``More than privacy: Applying
  differential privacy in key areas of artificial intelligence,'' \emph{IEEE
  Transactions on Knowledge and Data Engineering}, vol.~34, no.~6, pp.
  2824--2843, 2022.

\end{thebibliography}

\end{document}